\documentclass[review]{elsarticle}

\usepackage{hyperref}% add hypertext capabilities for linking the cross references
\usepackage{amsmath} 
\usepackage{graphicx}% Include figure files
\usepackage{bm}% bold math

\journal{arXiv.com}

\newcommand{\rd}{\mathrm d}

\begin{document}

\begin{frontmatter}

%%%%%%%%%%%%%%%%%%%%%%%%%%
\title{On the Room-Temperature QHE in Graphene}
%%%%%%%%%%%%%%%%%%%%%%%%%%
%
\author{S.~Fujita} 
\address{Department of Physics, University at Buffalo, State University of New York,\\ Buffalo,  NY~14260-1500, USA}
\author{A.~Suzuki%\footnote{Corresponding author: asuzuki@rs.kagu.tus.ac.jp}
}
\address{Department~of~Physics, Faculty~of~Science, Tokyo~University~of~Science,\\ Shinjyuku-ku, Tokyo~162-8601, Japan}

%%%%%%%%% Abstract
\begin{abstract}
The unusual quantum Hall effect (QHE) in graphene is described in terms of the composite (c-) bosons, which move with a linear dispersion relation.  The ``electron" (wave packet) moves easier in the direction $[1\,1\,0\,\,c\hbox{-axis}]\,\equiv\,[1\,1\,0]$ of the honeycomb lattice than perpendicular to it, while the ``hole" moves easier in $[0\,0\,1]$.  Since ``electrons" and ``holes" move in different channels, the particle densities can be high especially when the Fermi surface has ``necks".  The strong QHE arises from the phonon exchange attraction in the neighborhood of the ``neck"  surfaces.  The plateau observed for the Hall conductivity and the accompanied resistivity drop is due to the superconducting energy gap caused by the Bose-Einstein condensation of the c-bosons, each forming from a pair of one-electron--two-fluxons c-fermions by phonon-exchange attraction.  The half-integer quantization rule for the Hall conductivity: $\frac12(2P-1)(4e^2/h)$, $P=1, 2, ...$, is derived.
\end{abstract}
%%%%%%%
%
%\pacs{xxxx}
%

\end{frontmatter}

%\linenumbers

%% main text
%%%%%%%%%%%%%%%%%
\section{Introduction}\label{sec1}
\begin{figure}[tp]
\begin{center}
\includegraphics[scale=0.35]{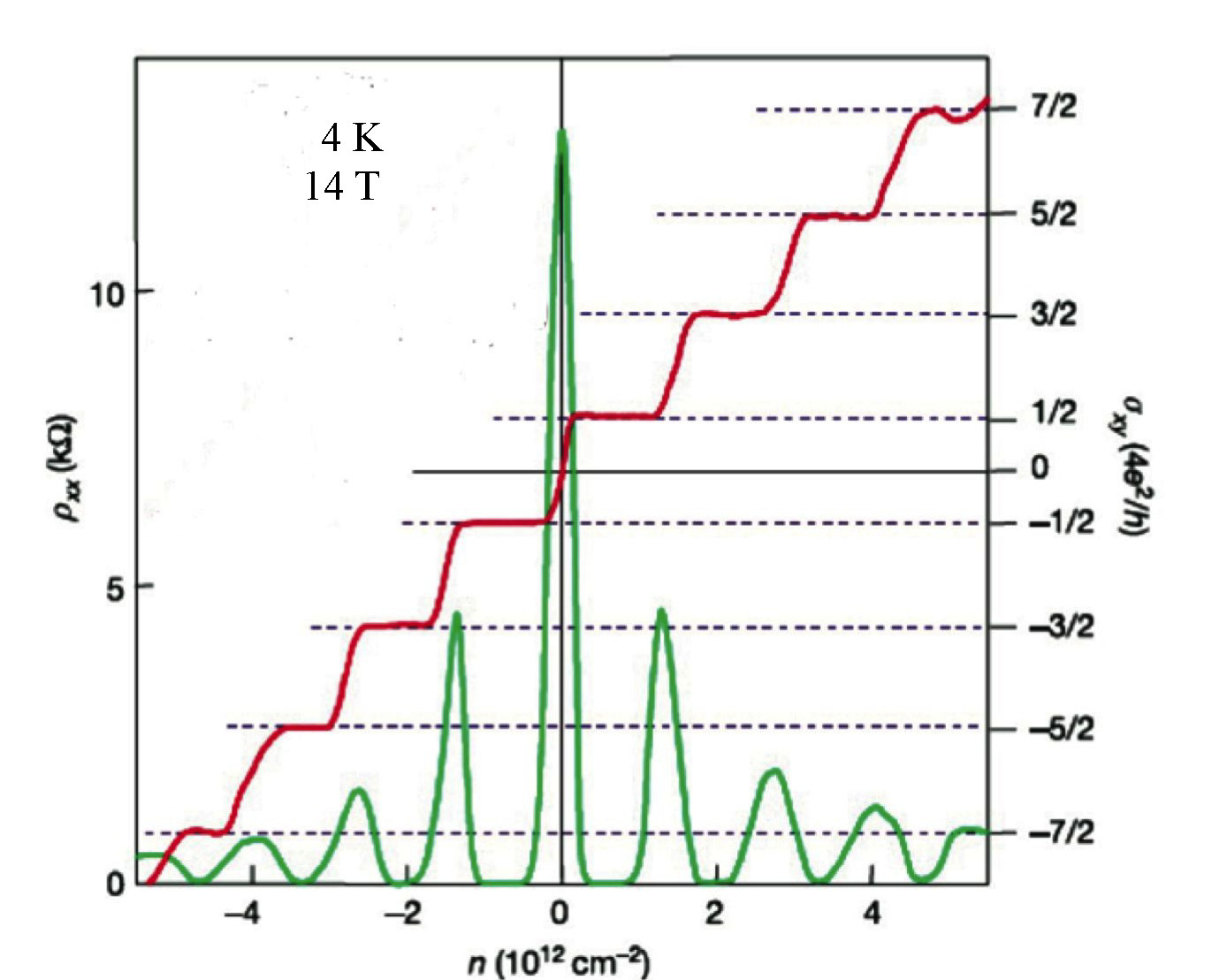}
%\centerline{\psfig{file=fig1.pdf,width=3.65in}}\vspace*{8pt}
\caption{(Color online) QHE in graphene.  Hall conductivity 
$\sigma_{xy}$ and longitudinal resistivity $\rho_{xx}$ are indicated by red and green lines, respectively, as a function of the conduction electron density.  After Novoselov {\it{et al.}}\cite{1}}
\label{fig1}
\end{center}
\end{figure}

%%%%%%%%%%%%%%%%%
%
In 2005 Novoselov {\it{et al.}}\,\cite{1} discovered a quantum Hall effect (QHE) in graphene, a single sheet of graphite.  Figure~\ref{fig1} is reproduced after Ref.\,\ref{1}, Fig.\,4.   
%
%%%%%Fig.1%%%%%
%%%%%%%%%%%%%%
%
The longitudinal magnetoresistivity $\rho_{xx}$ and the Hall conductivity $\sigma_{xy}$ in graphene at $B=14$\,T and $T=4$\,K are plotted as a function of the conduction electron (``electron" or ``hole") density $n$ in the scale of $10^{12}$\,cm$^{-2}$.
The plateau values of the Hall conductivity $\sigma_{xy}$ are quantized in the units of 
\begin{equation}%(1)
\frac{4e^2}{h}\label{1}
\end{equation}
within experimental errors, where $h$ is the Planck constant, $e$ the electron charge (magnitude).  The longitudinal resistivity $\rho_{xx}$ reaches zero at the middle of the plateaus.  These two are the major signatures of the QHE in graphene. 

\begin{figure}[tb]
\begin{center}
\includegraphics[scale=0.35]{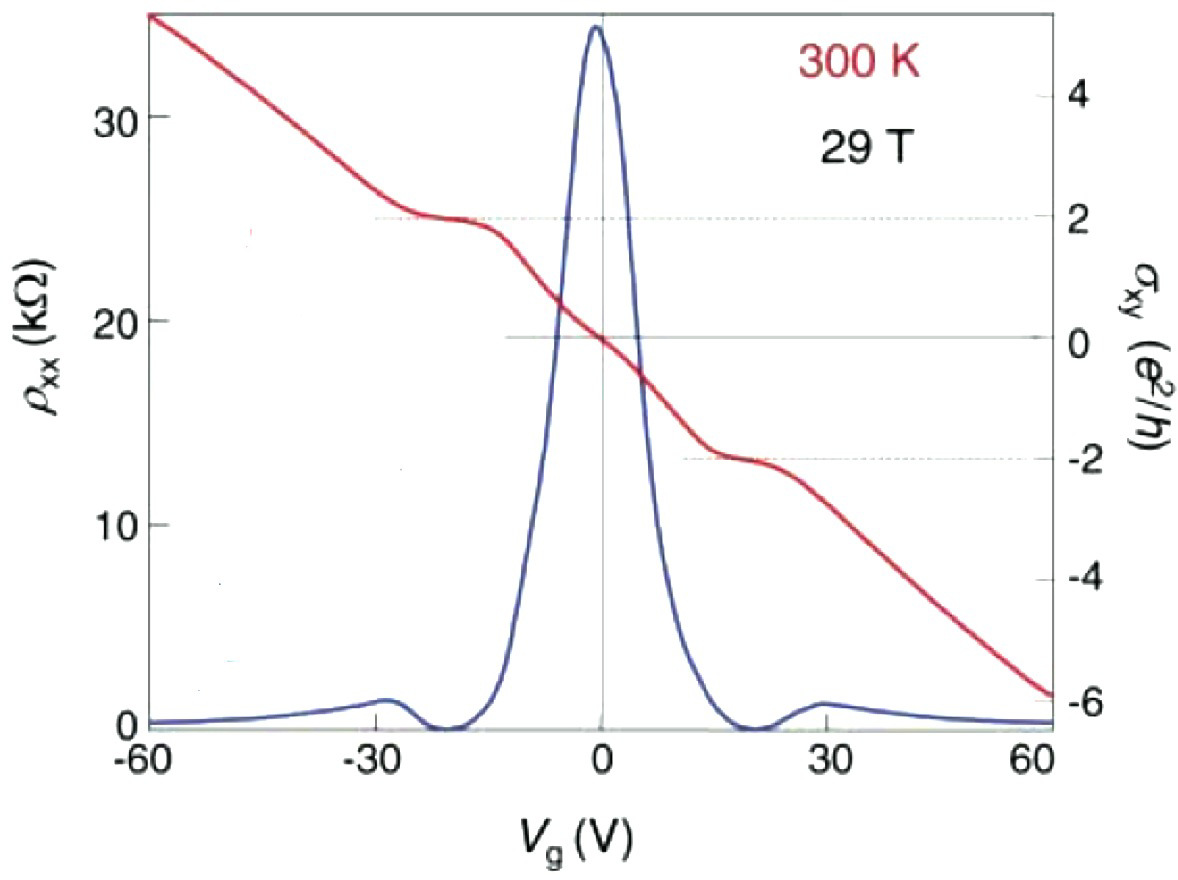}
%\centerline{\psfig{file=fig2.pdf,width=3.65in}}\vspace*{8pt}
\caption{(Color online) Room-temperature QHE in graphene after Novoselov {\it{et al.}}\,\cite{2}.  Hall conductivity $\sigma_{xy}(e^2/h)$ (red) and resistance $\rho_{xx}$ (blue) as a function of gate voltage ($V_{\mathrm g}$) at temperature 300\,K and magnetic field 29\,T.  Positive (negative) values of $V_{\mathrm g}$ indicate ``electrons'' (``holes'') at concentrations $n=(7.2 \times 10^{10}$ cm$^{-2}$V$^{-1}$) V$_{\mathrm g}$.}
\label{fig2}
\end{center}
\end{figure}

In 2007 Novoselov {\it{et al}}.\,\cite{2}  reported a discovery of a room temperature QHE in graphene.  We reproduceed  their data in  Figure\,\ref{fig2} after Ref.\,\ref{2}, Fig.\,1.  
%
%%%%%Fig.2%%%%%%%
%%%%%%%%%%%%%%%%%
%
The Hall resistivity $\rho_{xy}$ for ``electrons" and ``holes" indicate precise quantization within  experimental errors at magnetic field 29\,T and temperature 300\,K.  This is an extraordinary jump in the observation temperatures since the QHE in heterojunction GaAs/AlGaAs was reported below 0.5\,K.
Figure~\ref{fig2} is similar to those in Figure~\ref{fig1} although the abscissas are different, one in gate voltage and the other in carrier density, and hence the physical conditions are different.  We give an explanation later.  
Notice that the quantization in $\rho_{xy}$ appears in units of $h/4e^2$, which is a little strange since the most visible quantization for GaAs/AlGaAs appears in units of $h/e^2$.  We will resolve this mystery in the present work.

From the QHE behaviors in Figures\,\ref{fig1} and \ref{fig2}, 
we observe that the quantization in the Hall conductivity $\sigma_{xy}$ occurs at a set of half-integer points:
\begin{equation}%(2)
\frac{2P-1}{2}\left(\frac{4e^2}{h}\right)\,,\quad\,P=1,\,\,2,\,\,\cdots\,.\label{2}
\end{equation}
The original authors\,\cite{1,2} interpreted their data in terms of Dirac fermions.  A great number of experimental and theoretical papers followed.  The present work deals specifically with the quantization rule in Eq.\,(\ref{2}).   
We shall show this quantization rule based on the c-particles (fermions, bosons)\,\cite{PTEP} model in the present work.  We will defer discussion of Dirac fermions and the related matter. The preliminary results were reported in the conference proceedings\,\cite{3a}.
%
%%%%%%%%%%%%%%%%%%%%%
\section{Electron Dynamics in Graphene}\label{sec2}
%%%%%%%%%%%%%%%%%%%%%
%
The normal carriers in solids are ``electrons" (``holes''), which spiral around the applied magnetic field ${\bm B}$ counterclockwise (clockwise) viewed from the tip of the field vector ${\bm B}$.  The ``electrons'' (``holes'') are excited above (below) the metal's Fermi energy.  These quasiparticles are quotation marked throughout the text.  Following Ashcroft and 
Mermin\,\cite{3} we regard the conduction electrons as {\it{wave packets}}.  

\begin{figure}[hbp]
\begin{center}
\includegraphics[scale=0.75]{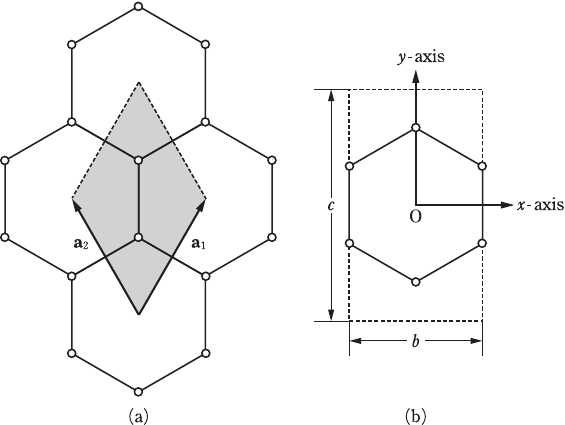}
%\centerline{\psfig{file=fig3.pdf,width=3.65in}}\vspace*{8pt}
\caption{(a) WS unit cell, rhombus (shaded) for graphene.  (b) The orthogonal unit cell, rectangle (dotted lines).}
\label{fig3}
\end{center}
\end{figure}

We consider a graphene, which forms a 2D honeycomb lattice.  The Wigner-Seitz (WS) unit 
cell\,\cite{4}, rhombus (shaded) shown in Figure\,\ref{fig3}\,(a), contains two C's.  We showed in our earlier work\,\cite{5} that graphene has ``electrons'' and ``holes'' based on the rectangular unit cell (dotted lines) shown in Figure~\ref{fig3}\,(b).  We briefly review our calculations.  We must choose the rectangular unit cell to establish the Bloch plane waves\,\cite{7} in 2D.  For a 1D space, there always exists a 1D $k$-space.  If one introduces non-orthogonal axes $(x_1,x_2)$ along $({\mathbf a}_1,{\mathbf a}_2)$, then one cannot use Fourier transformation.  This difficulty was discussed earlier in our previous work\cite{6}.  To establish the electron dynamics we need the orthogonal rectangular unit cell shown in Fig.\,\ref{fig3}\,(b).
%
%%%%%Fig. 3%%%%%%%
%%%%%%%%%%%%%%%%%
%

We assume that the ``electron'' (``hole'') wave packet has the charge $-e$ ($+e$) and a size of the rectangular unit cell, generated above (below) the Fermi energy $\varepsilon_{\mathrm F}$.  We showed\,\cite{5} that (a) the ``electron'' and the ``hole'' have different charge distributions and different effective masses, (b) that the ``electrons'' and ``holes'' move in different easy channels, (c) that the ``electrons'' and ``holes'' are thermally excited with different activation energies, and (d) that the ``electron'' activation energy $\varepsilon_1$ is smaller than the ``hole'' activation energy $\varepsilon_2$:
\begin{equation}%(3)
\varepsilon_1\,<\,\varepsilon_2\,.\label{3}
\end{equation}
The thermally activated electron densities are then given by
\begin{equation}%(4)
n_j(T)=n_je^{-\varepsilon_j/k_{\mathrm B}T}\,,\quad n_j=\hbox{constant},\label{4}
\end{equation}
where $j=1$ and 2 represent the ``electron'' and ``hole'', respectively.  In view of Eqs.\,(\ref{3}) and (\ref{4}), $n_1(T)>n_2(T)$.  Hence the ``electrons'' are the majority carriers in graphene.  Magnetotransport experiments by Zhang {\it{et al.}}\,\cite{9} indicate that the ``electrons'' are the majority carriers in graphene in agreement with experiments.
%
%%%%%%%%%%%%%%%%%%%
\section{Fractional Quantum Hall Effect}\label{sec3}
%%%%%%%%%%%%%%%%%%%
%
Fractional QHE were discovered by Tsui, Stormer and Gossard in 1982\,\cite{Tsui}.   In 1983 Laughlin proposed a revolutionary idea\,\cite{17} that fractional charges are carried by the elementary excitations for the fractional QHE system.  A great number of papers were 
followed\,\cite{18,22,16,19,15,20}.  Ezawa wrote books with extensive references for students and researchers\,\cite{14}.
The prevalent theories\,\cite{17,18,22,16,19,15,20} based on the Laughlin wave function\,\cite{17} in the Schr\"odinger picture deal with the QHE at 0 K and immediately above.  The system ground state, however, cannot carry a current.  To interpret the experimental data it is convenient to introduce composite (c-) particles (bosons, fermions).  The c-boson (c-fermion), each containing an electron and an odd (even) number of {\it magnetic flux quanta (fluxons)}, were introduced by Zhang {\it{et al.}}\,\cite{22} and others (Jain\,\cite{16}) for the description of the fractional QHE (Fermi liquid).  
The c-particles will be regarded as quasiparticles (elementary excitations) existing in the system.   A classical electron spirals around the applied static magnetic field.  The state has a lower energy relative to the original electron energy because the spiraling current (vortex) is diamagnetic.  The field-dressed (-attached) electron moves straight.  Jain\,\cite{16} established a close connection between the integer and the fractional QHE by introducing 
c-fermions.  His c-fermions are essentially the same as our c-fermions.  The types of mechanics (classical or quantum) do not change the energy sign.  A c-fermion is in a negative energy (bound) state.  Fujita and Okamura\,\cite{21} discussed the formation of a bound c-fermion and its connection with Jain's c-fermion.  Jain did not include the c-bosons in his book\,\cite{Jain}.  We view the 
c-bosons as equally important as the c-fermions.  A c-boson is also in a bound state.  Besides, c-bosons can be Bose-Einstein (BE) condensed, which generates a stabilizing (superconducting) energy gap in the excitation spectrum.   All QHE states with distinctive Hall plateaus in heterojunction GaAs/AlGaAs are observed below the critical temperature $T_{\mathrm c}\simeq 0.5$\,K.  The QHE in graphene observed at 300\,K is an exception.  It is desirable to treat the QHE below and above $T_{\mathrm c}$ in a unified manner.  The extreme accuracy (precision $\sim\,10^{-8}$) in which each Hall plateau is observed means that the current density $j$ must be computed exactly without averaging.  In the prevalent theories\,\cite{17,18,22,16,19,15,20}, the electron-electron interaction and Pauli's exclusion principle are regarded as the cause for the QHE.  Both are essentially repulsive and cannot account for the fact that the c-particles are bound, that is, they are in negative-energy states.  Besides, the prevalent theories have limitations:
\begin{itemize}
\item The zero temperature limit is taken at the outset.  Then the question why QHE is observed below 0.5\,K in GaAs/AlGaAs cannot be answered.  We better have a theory for all temperatures.
\item The high-field limit is taken at the outset.  The integer QHE at filling factor (Landau level occupation number) $\nu=P$ are observed for small integer $P$ only.  The question why the QHE for high $P$ (weak field)  is not observed cannot be answered.  We better describe the phenomena for all fields.
\item The Hall resistivity $\rho_{\mathrm H}$ value $(Q/P)(h/e^2)$ is obtained in a single stroke.  To obtain $\rho_{\mathrm H}$ we need two separate measurements of the Hall field $E_{\mathrm H}$ and the current density $j$.  We must calculate $(E_{\mathrm H}, j)$ and take the ratio $E_{\mathrm H}/j$ to obtain $\rho_{\mathrm H}$.  
\end{itemize}

Fujita and Okamura\,\cite{21} developed a quantum statistical theory based on phonon exchange attraction, and used Laughlin's results to describe the fractional QHE.  In the present work we complete the description without using Laughlin's fractional charge idea with the assumtion that any c-fermion has the charge magnitude $e$.   See the paper by Fujita, Suzuki and Ho\,\cite{FSH} for more detail.  There is a remarkable similarity between the QHE {\it{and}} the High-Temperature Superconductivity (HTSC), both occurring in 2D systems as pointed out by Laughlin\,\cite{17a}.  We regard the {\it{phonon exchange attraction}} as the causes of both QHE and superconductivity.  Starting with a reasonable Hamiltonian, we calculate everything using the standard statistical mechanics.  

The countability concept of the fluxons, known as the {\it{flux quantization}}:
\begin{equation}%(5)
B=\frac{N_\phi}{A}\frac{h}{e}\equiv n_\phi\Phi_0\,,\quad n_\phi \equiv \frac{N_\phi}{A}\,,\label{5}
\end{equation}
where $A=$ sample area, $N_\phi=$ fluxon number (integer), $\Phi_0\equiv h/e=$ flux quantum, is originally due to Onsager\,\cite{23}.  The magnetic (electric) field is an axial (polar) vector and the associated fluxon (photon) is a half-spin fermion (full-spin boson).  The magnetic (electric) flux line cannot (can) terminate at a sink, which supports the fermionic (bosonic) nature of the associated fluxon (photon).  No half-spin fermion can annihilate itself because of angular momentum conservation.  The electron spin originates in the relativistic electron equation (Dirac's theory of electron)\,\cite{24}.  The discrete (two) quantum numbers $(\sigma_z=\pm1)$ cannot change in the continuous limit, and hence the spin must be conserved.  The countability and statistics of the fluxon is the fundamental particle properties.  We postulate that  {\it the fluxon is a half-spin fermion with zero mass and zero charge}.  

We assume that the magnetic field ${\bm B}$ is applied perpendicular to the 2D plane.  The 2D Landau level energy,
\begin{equation}%(6)
\varepsilon=\hbar\omega_{\mathrm c}\left(N_{\mathrm L}+\frac12\right)\,,\quad\omega_{\mathrm c}\equiv eB/m^*\,,\quad N_{\mathrm L}=0,1,2,\cdots \label{6}
\end{equation}
with the states $(N_{\mathrm L}, k_y)$ have a great degeneracy; the $m^*$ is the effective mass of an ``electron'' and the $\omega_{\mathrm c}$ the cyclotron frequency.  The Center-of-Mass (CM) of {\it{any}} c-particle moves as a fermion (boson).  The eigenvalues of the CM momentum are limited to 0 or 1 (unlimited) if it contains an odd (even) number of elementary fermions.  This rule is known as the {\it{Ehrenfest-Oppenheimer-Bethe's}} (EOB's) {\it{rule}}\,\cite{24,25,26}.  Hence the CM motion of the composite containing an electron and  $Q$ fluxons is bosonic (fermionic) if $Q$ is odd (even).  The system of the c-bosons condenses below the critical temperature $T_{\rm c}$ and exhibits a superconducting state while the system of c-fermions shows a Fermi liquid behavior.

A longitudinal phonon, acoustic or optical, generates a density wave, which affects the electron (fluxon) motion through the charge displacement (current).  The exchange of a phonon between electron and fluxon generate an {\it{attractive}} transition.  

Bardeen, Cooper and Schrieffer (BCS)\,\cite{28} assumed the existence of Cooper pairs\,\cite{29} in a superconductor, and wrote down a Hamiltonian containing the ``electron'' and ``hole'' kinetic energies and the pairing interaction Hamiltonian with the phonon variables eliminated.  We start with a BCS-like Hamiltonian ${\mathcal H}$ for the QHE:\,\cite{21}
%
%\begin{widetext}
\begin{align}%(7)
{\mathcal H}=&{\sum_{\bm k}}^{\prime}\sum_{s}\varepsilon_{\bm k}^{(1)}n_{{\bm k}s}^{(1)}
+{\sum_{\bm k}}^{\prime}\sum_{s}\varepsilon_{\bm k}^{(2)}n_{{\bm k}s}^{(2)}+{\sum_{\bm k}}^{\prime}
\sum_{s}\varepsilon_{\bm k}^{(3)}n_{{\bm k}s}^{(3)}\nonumber\\
&-{\sum_{\bm q}}^{\prime}{\sum_{\bm k}}^{\prime}{\sum_{{\bm k}^{\prime}}}^{\prime}
\sum_{s}v_{0}\left[ B_{{\bm k}^{\prime}{\bm q}\,s}^{(1)\dagger}B_{{\bm k}{\bm q}\,s}^{(1)}+B_{{\bm k}^{\prime}{\bm q}\,s}^{(1)\dagger}B_{{\bm k}{\bm q}\,s}^{(2)\dagger}%\right.\nonumber\\&
+B_{{\bm k}^{\prime}{\bm q}\,s}^{(2)}B_{{\bm k}{\bm q}\,s}^{(1)}+B_{{\bm k}^{\prime}{\bm q}\,s}^{(2)}B_{{\bm k}{\bm q}\,s}^{(2)\dagger} \right],\label{7}
\end{align}
%\end{widetext}
%
where $n_{{\bm k}s}^{(j)}=c_{{\bm k}s}^{(j)\dagger}c_{{\bm k}s}^{(j)}$ is the number operator for the ``electron'' (1) [``hole''\,(2), fluxon\,(3)] at momentum ${\bm k}$ and spin $s$ with the energy $\varepsilon_{\bm k}^{(j)}$ with annihilation (creation) operators $c$ 
$(c^\dagger)$ satisfying the Fermi anticommutation rules:
\begin{align}%(8)
\{c_{{\bm k}s}^{(i)},\,c_{{\bm k}^{\prime}s^{\prime}}^{(j)\dagger}\}
&\equiv  c_{{\bm k}s}^{(i)}c_{{\bm k}^{\prime}s^{\prime}}^{(j)\dagger} + c_{{\bm k}^{\prime}s^{\prime}}^{(j)\dagger}c_{{\bm k}s}^{(i)}=\delta_{{\bm k},{\bm k}^{\prime}}\delta_{s,s^{\prime}}\delta_{i,j}\,,\quad%\nonumber\\
\{c_{{\bm k}s}^{(i)}, c_{{\bm k}^{\prime}s^{\prime}}^{(j)}\}=0\,.\label{8}
\end{align}
The fluxon number operator $n_{{\bm k}s}^{(3)}$ is represented by $a_{{\bm k}s}^\dagger a_{{\bm k}s}$ with $a$ $(a^\dagger)$ satisfying  the anticommutation rules:
\begin{equation}%(9)
\{a_{{\bm k}s},\,a_{{\bm k}^{\prime}s^{\prime}}^\dagger\} = \delta_{{\bm k},{\bm k}^{\prime}}\delta_{s,s^{\prime}}\,,\quad\{a_{{\bm k}s},\,a_{{\bm k}^{\prime}s^{\prime}}\}=0\,.\label{9}
\end{equation}

The phonon exchange can create electron-fluxon composites, bosonic or fermionic, depending on the number of fluxons.  We call the conduction-electron composite with an odd (even) number of fluxons {\it{c-boson}}\,({\it{c-fermion}}).  The electron\,(hole)-type c-particles carry negative (positive) charge.  Electron\,(hole)-type Cooper-pair-like c-bosons are generated by the phonon-exchange attraction from a pair of electron (hole)-type c-fermions.  The pair operators $B$ are defined by
 \begin{align}%(10)
 B_{{\bm{kq}},s}^{(1)\dagger}&\equiv c_{{\bm k}+{\bm q}/2,s}^{(1)\dagger}c_{-{\bm k}+{\bm q}/2,-s}^{(1)\dagger}\quad\text{for ``electrons"},\nonumber\\
 B_{{\bm{kq}},s}^{(2)}&\equiv c_{-{\bm k}+{\bm q}/2,-s}^{(2)}c_{{\bm k}+{\bm q}/2,s}^{(2)}\quad\text{for ``holes"}.\label{10}
 \end{align}
The prime on the summation in Eq.\,(\ref{7}) means the restriction: $0<\varepsilon_{{\bm k}s}^{(j)}<\hbar\omega_{\rm D}$, $\omega_{\rm D}=$ Debye frequency.    
The pairing interaction terms in Eq.\,(\ref{7}) conserve the charge.  The term $-v_{0}B_{{\bm k}^{\prime}{\bm q}\,s}^{(1)\dagger}B_{{\bm k}{\bm q}\,s}^{(1)}$, where $v_{0}\equiv \vert V_{{\bm q}}V_{{\bm q}}^{\prime}\vert\,(\hbar\omega_{0}A)^{-1}$, $A=$ sample area, is the pairing strength, generates a transition in the electron-type c-fermion states.    Similarly, the exchange of a phonon generates a transition between the hole-type c-fermion states,  represented by $-v_{0}B_{{\bm k}^{\prime}{\bm q}\,s}^{(2)\dagger}B_{{\bm k}{\bm q}\,s}^{(2)\dagger}$.  The phonon exchange can also pair-create\,(pair-annihilate) electron\,(hole)-type c-boson pairs, and the effects of these processes are represented by $-v_{0}B_{{\bm k}^{\prime}{\bm q}\,s}^{(1)\dagger}B_{{\bm k}{\bm q}\,s}^{(2)\dagger}$ $(-v_{0}B_{{\bm k}{\bm q}\,s}^{(1)}B_{{\bm k}{\bm q}\,s}^{(2)})$.  

The Cooper pair is formed from two ``electrons'' (or ``holes'').  Likewise the c-bosons may be formed by the phonon-exchange attaraction from two like-charge c-fermions.  If the density of the c-bosons is high enough, then the c-bosons will be BE-condensed and exhibit a superconductivity.

The pairing interaction terms in Eq.\,(\ref{7}) are formally identical with those in the generalized BCS Hamiltonian\,\cite{30}.    
Only we deal here with c-fermions instead of conduction electrons.  

The c-bosons, having the linear dispersion relation, can move in all directions in the plane with the constant speed $(2/\pi)v_{\mathrm F}^{(j)}$~\cite{21,30}.  The supercurrent is generated by $\mp$ c-bosons monochromatically condensed, running along the sample length.  The supercurrent density (magnitude) $j$, calculated by the rule: $j=(\hbox{carrier charge}\!: e^*)\times(\hbox{carrier density}\!: n_0)\times(\hbox{carrier drift velocity}\!: v_{\mathrm d})$, is given by
\begin{equation}%(11)
j\equiv e^*n_0v_{\mathrm d}=e^*n_0\,\frac{2}{\pi}\left\vert v_{\mathrm F}^{(1)} - v_{\mathrm F}^{(2)}\right\vert\,,\label{11}
\end{equation}
where $e^{*}$ is the {\it{effective}} charge of carriers.  
The Hall field (magnitude) $E_{\mathrm H}$ equals $v_{\mathrm d}B$.  The magnetic flux is quantized as in Eq.\,(\ref{5}).  Hence we obtain
\begin{equation}%(12)
\rho_{\rm H}\equiv\frac{E_{\rm H}}{j}=\frac{v_{\rm d}B}{e^*n_0v_{\rm d}}
=\frac{1}{e^*n_0}n_{\phi}\Phi_0
\equiv\frac{n_{\phi}}{e^*n_0}\left(\frac{h}{e}\right)\,.
\label{12}
\end{equation}
Here, we assumed that {\it the c-fermion containing an electron and an even number of fluxons} has a charge magnitude $e$.
For the integer QHE, $e^*=e$, $n_{\phi}=n_0$, then we obtain $\rho_{\mathrm H}=h/e^2$, explaining the plateau value observed for the integer QHE.

The supercurrent generated by equal numbers of $\mp$ c-bosons condensed monochromatically is neutral.  This is reflected in the calculations in Eq.\,(\ref{11}).  The supercondensate whose motion generates the supercurrent must be neutral.  If it has a charge, it would be accelerated indefinitely by the external field because the impurities and phonons cannot stop the supercurrent to grow.  That is, the circuit containing a superconducting sample and a battery must be burnt out if the supercondensate is not neutral.  In the calculation of $\rho_{\mathrm H}$ in Eq.\,(\ref{12}), we used the unaveraged drift velocity 
$v_{\mathrm d}=(2/\pi)\vert v_{\mathrm F}^{(1)}-v_{\mathrm F}^{(2)}\vert$, which is significant.  Only the unaveraged drift velocity cancels out exactly from numerator/denominator, leading to an exceedingly accurate plateau value.  

We now extend our theory to include elementary fermions (electron, fluxon) as members of the c-fermion set.  We can then treat the superconductivity and the QHE in a unified manner.  The c-boson containing one electron and one fluxon can be used to describe the integer QHE.

Important pairings and the effects are listed below.
\begin{itemize}
\item a pair of conduction electrons,  superconductivity
\item a fluxon and c-fermions,  QHE  
\item a pair of like-charge conduction electrons, each with two fluxons,  QHE in graphene.
\end{itemize}
%
%%%%%%%%%%%%%%%%%%%
\section{The Room Temperature QHE}\label{sec4}
%%%%%%%%%%%%%%%%%%%
%
The QHE behavior observed for graphene is remarkably similar to that for GaAs/AlGaAs.  The physical conditions are different however since the gate voltage and the applied magnetic field are varied in the experiments.  The present authors regard the QHE in GaAs/AlGaAs as a manifestation of superconductivity generated by the magnetic field.  Briefly, the magnetoresistivity for a QH system reaches zero (superconducting) and the accompanied Hall resistivity generates a plateau by the Meissner effect.  The QHE state is not easy to destroy because of the {\it superconducting energy gap} in the c-boson excitation spectrum.  If an extra magnetic field is applied to the system at optimum QHE state (the center of the plateau), then the system remains in the same superconducting state by expelling the extra field.  If the field is reduced, then the system stays in the same state by sucking in extra field fluxes, thus generating a Hall conductivity plateau.  In the graphene experiments, the gate voltage is varied.  A little extra gate voltage relative to the optimum voltage (the center of the plateau) polarizes the system without changing the superconducting state, thus generating a Hall conductivity plateau.  This state has an extra electric field energy:
\begin{equation}%(13)
\frac A2\varepsilon_{0}(\Delta E)^{2}\,, \label{13}
\end{equation}
where $A$ is the sample area, $\varepsilon_{0}$ the dielectric constant, and $\Delta E$ is the extra electric field, positive or negative, depending on the field direction.  If the gate voltage is further increased (or decreased), then it will eventually destroy the superconducting state, and the resistivity will rise from zero.  A strong current generates high magnetic field around it, which eventually destroys the supercurrent.  This explains the flat $\sigma_{xy}$ plateau and the rise in resistivity from zero.

We now examine the data shown in {Figure~\ref{2}}.  We first  observe that the right-left symmetry is broken.  ``Electrons'' and ``holes'' move in different channels with different masses, breaking symmetry.  The applied gate voltage induce the surface conduction electrons and hence changes the Fermi surface.  A relatively high voltage 20\,V may bring the system to the van Hove singularity points in the neighborhood of which the conduction electron densities are high.  This is where the prominent QHE is observed.  We note that such discussions are possible only with the rectangular unit cell model, and not with the WS unit cell model, which predicts a gapless semiconductor with the electron-hole symmetry: $m_1=m_2$, $\varepsilon_1=\varepsilon_2$.

We wish to derive the quantization rule in Eq.\,(\ref{2}).  Let us first consider the case of $P=1$.  
The QHE  requires a BEC of c-bosons.  Its favorable environment is near the van Hove singularities, where the Fermi surface changes its curvature sign.  For graphene, this happens when the 2D Fermi surface just touches the Brillouin zone boundary and ``electrons" or ``holes" are abundantly generated.  The quantization rule given by Eq.\,(\ref{2}) is realized if the c-bosons are formed from a pair of like-charge c-fermions, each containing a conduction electron and two (2) fluxons.  
By assumption, {\it each c-fermion has the effective charge $e$}:
\begin{equation}%(14)
e^*=e \quad{\hbox{for any c-fermion.}}\label{14}
\end{equation}
After studying the low-field QH states of c-fermions we obtain
\begin{equation}%(15)
n_\phi^{(Q)}=n_{\mathrm e}/Q\,,\quad Q=0,\,2,\,4,\,\cdots\,,\label{15}
\end{equation}
for the density of the c-fermions with $Q$ fluxons, where $n_{\mathrm e}$ is the electron density.  All fermionic QH states (points) lie on the classical-Hall straight line passing the origin with a constant slope when $\sigma_{\mathrm H}$ is plotted as a function of the inverse magnetic field.  For higher fields the LL spacing $\hbar\omega_{\mathrm c}$ is greater, and hence the fermion formation is more difficult if $Q$ is greater.  The c-boson contains two (2) c-fermions.  Using Eq.\,(\ref{12}), we obtain
\begin{equation}%(16)
\sigma_{\mathrm H}\equiv\rho_{\mathrm H}^{-1}
=\frac{j}{E_{\mathrm H}}
=\frac{2en_0v_{\mathrm d}}{v_{\mathrm d}B}
=\frac{2en_0}{n_\phi\Phi_0}%=\frac{en_0}{n_\phi}\left(\frac{e}{h}\right)
=\frac{2e^2}{h}\,.\label{16}
\end{equation}
Here, the field $B=n_\phi\Phi_0$ at $\nu=1/2$ is used, where the c-boson density $n_0$ is equal to the flux density $n_\phi$.  We note that the value $2e^2/h$ obtained here is in agreement with the experiments shown in Fig.\,1.

\begin{figure}[htb]
\begin{center}
\includegraphics[scale=0.85]{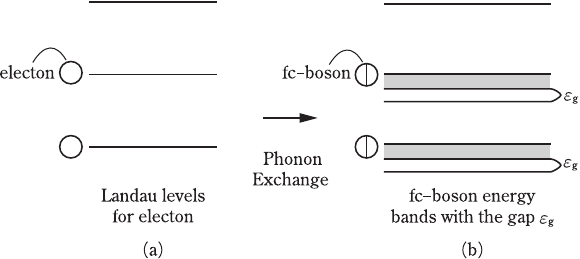}
%\centerline{\psfig{file=fig4.pdf,width=3.65in}}\vspace*{8pt}
\caption{The electrons which fill up the lowest two LL's, form the QH state at $\nu=2$ after the phonon-exchange attraction and the BEC of the c-bosons.}
\label{fig4}
\end{center}
\end{figure}

The QHE states with integers $P=1,\,2,\,\cdots$ are generated on the weaker field side.  Their strengths decrease with increasing $P$ as shown below.  The magnetic field magnitude becomes smaller with increasing $P$.  The LL degeneracy is proportional to $B$, and hence $P$ LL's must be considered.  First consider the case $P=2$.  Without the phonon-exchange attraction the electrons occupy the lowest two LL's with spin.  The electrons at each level form fundamental (f)\,c-bosons.   
In the superconducting state the c-bosons occupy the monochromatically condensed state, which is separated by the superconducting gap $\varepsilon_{\mathrm g}$ from the continuum states (band) as shown in the right-hand figure in Fig.\,\ref{fig4}.
%
%%%%%%%%%%%%%% Fig.4 %%%%%%%%%%%%%
%%%%%%%%%%%%%%%%%%%%%%%%%%%%%%%
%
The c-boson density $n_0$ at each LL is one-half the density at $\nu=1$, which is equal to the 
electron density $n_{\mathrm e}$ fixed for the sample.  Extending the theory to a general integer $P$, we have
\begin{equation}%(24)
n_0=n_{\mathrm e}/P\,.\label{24}
\end{equation}
This means that both the critical temperature $T_{\mathrm c}\,(\propto n_0^{1/2})$ and the energy gap $\varepsilon_{\mathrm g}$ are smaller, making the plateau width (a measure of $\varepsilon_{\mathrm g}$) smaller in agreement with experiments.  The c-bosons have lower energies than the conduction electrons.  Hence at the extreme low temperatures the supercurrent due to the condensed c-bosons dominates the normal currents due to the conduction electrons and non-condensed c-bosons, giving rise to the dip in $\rho$.  The superconducting energy gap $\varepsilon_{\mathrm g}(T)$ is obtained and discussed earlier.  For completeness the derivation of $\varepsilon_{\mathrm g}$ is given in Appendix.
Thus, we have obtained Eq.\,(\ref{2}) within the framework of our fractional QHE theory in terms of c-particles.  

In summary, we established that
\begin{itemize}
\item The half-integer FQHE arises from the BEC of c-bosons, each containing a pair of c-fermions with two fluxons.
\item The Hall conductivity $\sigma_{xy}$ is quantized at
$\frac12(2P-1)\left({4e^2}/{h}\right)$, $P=1,\,2,\,\cdots$.
\item The strengths of the plateaus become smaller with increasing $P$.
\end{itemize}
%
%%%%%%%%%%%%%%%%%%%%%%%%%%%%%%%%%%%%
\section*{Appendix:  Temperature Dependent Energy Gap $\varepsilon_{\mathrm g}(T)$}
%%%%%%%%%%%%%%%%%%%%%%%%%%%%%%%%%%%%
%
The c-bosons can be bound by the interaction Hamiltonian $-v_0B_{{\bm k}^\prime{\bm q}}^{(j)\dagger}B_{{\bm k}{\bm q}}^{(j)}$.  The fundamental c-bosons (fc-bosons) can undergo a Bose-Einstein condensation (BEC) below the critical temperature {$T_{\mathrm c}$}.  The fc-bosons are condensed at a momentum along the sample length.  Above $T_{\mathrm c}$, they can move in all directions in the plane with the Fermi speed $v_{\mathrm F}^{(j)}$.  The ground state energy $w_0$ can be calculated by solving the Cooper-like equation:
\begin{equation}%(A1)
w_0\Psi({\bm k})=\varepsilon_{\bm k}\Psi({\bm k})-\frac{v_0}{(2\pi\hbar)^2}\int^\prime\rd^2k^\prime\Psi({\bm k}^\prime)\,,\tag{A1}
\end{equation}
where $\Psi$ is the reduced wave function for the stationary fc-bosons; the prime on the integral sign means that the restriction: $0<\varepsilon_k<\hbar\omega_{\mathrm D}$, $\omega_{\mathrm D}$=Debye frequency.  We obtain after simple calculations
\begin{equation}%(A2)
w_0=\frac{-\hbar\omega_{\mathrm D}}{\exp\left\{1/(v_0{\mathcal D}_0)\right\}- 1}\,<\,0\,, \tag{A2}
\end{equation}
where ${\mathcal D}_0\equiv{\mathcal D}(\varepsilon_{\mathrm F})$ is the density of states per spin at $\varepsilon_{\mathrm F}$.  Note that the binding energy $\vert w_0\vert$ does not depend on the ``electron" mass.  Hence, the $\pm$fc-bosons have the same energy $w_0$.

At 0 K only {\it stationary} fc-bosons are generated.  The ground state energy $W_0$ of the system of fc-bosons is
\begin{equation}%(A3)
W_0=2N_0w_0\,,\tag{A3}
\end{equation}
where $N_0$ is the $-$ (or $+$) fc-boson number.

At a finite $T$ there are moving (non-condensed) fc-bosons, whose energies $w_{\bm q}^{(j)}$ are obtained from\cite{31}
\begin{equation}%(A4)
w_{\bm q}^{(j)}\Psi({\bm k},{\bm q})=\varepsilon^{(j)}_{\vert{\bm k}+{\bm q}\vert}\Psi({\bm  k},{\bm q})-\frac{v_{0}}{(2\pi\hbar)^{2}}\int^{\prime}\rd^{2}k^{\prime}\Psi({\bm k}^{\prime},{\bm q})\,.\tag{A4}
\end{equation}
For small $q$, we obtain
\begin{equation}%(A5)
w_{q}^{(j)}=w_{0}+\frac{2}{\pi}v_{\mathrm F}^{(j)}\vert{\bm q}\vert\,,\tag{A5}
\end{equation}
where $v_{\mathrm F}^{(j)}\equiv(2\varepsilon_{\mathrm F}/m_j)^{1/2}$ is the Fermi speed.  The energy $w_q^{(j)}$ depends {\it linearly} on the momentum magnitude {$q$}.  

The system of free massless bosons undergoes a BEC in 2D at the critical temperature $T_{\mathrm c}$:
\begin{equation}%(A6)
k_{\mathrm B}T_{\mathrm c}=1.945\,\hbar cn^{1/2}\,,\tag{A6}
\end{equation}
where $c$ is the boson speed, and $n$ the density.  Briefly the BEC occurs when the chemical potential $\mu$ vanishes at a finite $T$.  The critical temperature $T_{\mathrm c}$ can be determined from
 \begin{equation}%(A7)
 n=(2\pi\hbar)^{-2}\int \rd^{2}p\,[e^{\beta_{\mathrm c}\varepsilon}-1]^{-1}\,,\quad\beta_{\mathrm c}\equiv(k_{\mathrm B}T_{\rm c})^{-1}\,.\tag{A7}
 \end{equation}
 After expanding the integrand in powers of $e^{-\beta_{\mathrm c}\varepsilon}$ and using $\varepsilon=cp$, we obtain
 \begin{equation}%(A8)
 n=1.654\,(2\pi)^{-1}(k_{\mathrm B}T_{\mathrm c}/\hbar c)^2\,,\tag{A8}
 \end{equation}
from which we obtain formula (A6).  Substituting $c=(2/\pi)v_{\mathrm F}$ in Eq.\,(A6), we obtain
\begin{equation}%(A9)
k_{\mathrm B}T_{\mathrm c}=1.24\,\hbar v_{\mathrm F}n_0^{1/2}\,,\qquad n_0\equiv {N_0/A\,.}\tag{A9}
\end{equation}
The interboson distance $R_0\equiv1/\sqrt{n_0}$ calculated from this equation is $1.24 \hbar v_{\mathrm F}/(k_{\mathrm B}T_{\mathrm c})$.  The boson size $r_0$ calculated from Eq.\,(A9), using the uncertainty relation $(q_{\mathrm{max}}r_0$ $\sim\hbar)$ and $\vert w_0 \vert\,\sim\,k_{\mathrm B}T_{\mathrm c}$, is $r_0=(2/\pi)\hbar v_{\mathrm F}(k_{\mathrm B}T_{\mathrm c})^{-1}$, which is a few times smaller than $R_0$.  Thus the bosons do {\it not} overlap in space, and the {\it free boson model} is justified.

In the presence of the BE-condensate below ${T}_{\mathrm c}$,  the unfluxed electron carries the energy $E_{{\bm k}}^{(j)}=(\varepsilon_{\bm k}^{(j)2} + {\Delta}^{2})^{1/2}$, where the quasielectron energy gap {$\Delta$} is the solution of
\begin{equation}%(A10)
1=v_0{\mathcal D}_0\int_0^{\hbar\omega_{\mathrm D}}\rd\varepsilon\frac{1}{(\varepsilon^{2}+\Delta^{2})^{1/2}}\Big\{1+\exp[-\beta(\varepsilon^{2}+\Delta^{2})^{1/2}]\Big\}^{-1}\,,\quad\beta\equiv(k_{\mathrm B}T)^{-1}\,.\tag{A10}
\end{equation}
Note that the gap $\Delta$ depends on $T$.  At $T_{\mathrm c}$ there is no condensate, and hence $\Delta$ vanishes.

The {\it moving} fc-boson below $T_{\mathrm c}$ with the condensate background has the energy $\widetilde w_{\bm q}$, obtained from
\begin{equation}%(A11)
\widetilde w^{(j)}_{\bm q}\Psi({\bm k},{\bm q})=E_{\vert{\bm k}+{\bm q}\vert}^{(j)}\Psi({\bm k},{\bm q})-\frac{v_0}{(2\pi\hbar)^2}\int^{\prime}\rd^{2}k^{\prime}\Psi({\bm k}^{\prime},{\bm q})\,,\tag{A11}
\end{equation}
where $E^{(j)}$ replaced $\varepsilon^{(j)}$ in Eq.\,(A4).  We obtain
\begin{equation}%(A12)
\widetilde w_{\bm q}^{(j)}=\widetilde w_{0}+\frac{2}{\pi}v_{\mathrm F}^{(j)}\vert{\bm q}\vert=w_{0}+\varepsilon_{\mathrm g}+\frac{2}{\pi}v_{\mathrm F}^{(j)}q\,,\tag{A12}
\end{equation}
where ${\widetilde w}_0(T)$ is determined from
\begin{equation}%(A13)
1={\mathcal D}_{0}v_{0}\int_{0}^{\hbar\omega_{\mathrm D}}\frac{\rd\varepsilon}{\vert\widetilde w_{0}\vert + (\varepsilon^{2}+\Delta^{2})^{1/2}}\,. \tag{A13}
\end{equation}
The energy difference
\begin{equation}%(A14)
\widetilde w_{0}(T)-w_{0}\equiv\varepsilon_{\mathrm g}(T) > 0 \tag{A14}
\end{equation}
represents the $T$-{\it dependent energy gap} between the moving and stationary fc-bosons.  The energy $\widetilde w_{\bm q}$ is negative.  Otherwise, the fc-boson should break up.  This limits $\varepsilon_{\mathrm g}$ to be less than $\vert w_{0}\vert$.  The energy gap $\varepsilon_{\mathrm g}(T)$ is $\vert w_{0}\vert$ at 0 K.  
It declines to zero as the temperature approaches ${T}_{\mathrm c}$.

The experimental electron density is $3.16\times 10^{12}$ cm$^{-2}$ and  the Fermi velocity $v_{\mathrm F}=1.1\times10^6$ ms$^{-1}$.  The critical temperature $T_{\mathrm c}$ is expected to be much above 300 K.  The temperature 50 K can be regarded as a very low temperature relative to $T_{\mathrm c}$.  Hence the QH state has an Arhenius-decay type exponential stability factor:
\begin{equation}%(A15)
\exp[-\varepsilon_{\mathrm g}(T=0)/k_{\mathrm B}T]\,, \tag{A15}
\end{equation}
where {$\varepsilon_{\mathrm g}(T=0)$} is the zero-temperature energy gap.
%
%\section*{References}
%%%%%%%%%%%%%%%

\end{document}